\documentclass[11pt, A4]{article}

\usepackage{graphicx}
\usepackage{epstopdf}
\usepackage{amsmath}
\usepackage{amssymb}
\usepackage{amsfonts}
\usepackage{amsthm}
\usepackage[usenames]{color}
\usepackage{array}

\usepackage[
      colorlinks=true,
      linkcolor=blue,
      urlcolor=blue,
      filecolor=blue,
      citecolor=red,
      pdfstartview=FitV,
      pdftitle={},
      pdfauthor={},
      pdfsubject={},
      pdfkeywords={},
      pdfpagemode=None,
      bookmarksopen=true
]{hyperref}

\usepackage{epsfig}
\usepackage{hyperref}

\def\cL{\mathcal{L}}
\def\cM{\mathcal{M}}

\def\cT{\mathcal{T}}

\def\cV{\mathcal{V}}

\def\beq{\begin{eqnarray}}
\def\eeq{\end{eqnarray}}

\def\mf{\mathfrak}
\def\s{\sigma}

\def\be{\begin{equation}}
\def\ee{\end{equation}}
\def\bea{\begin{eqnarray}}
\def\eea{\end{eqnarray}}

\newcommand{\rom}[1]{\mathrm{#1}}

\def\cL{\mathcal{L}}
\def\cM{\mathcal{M}}

\def\cT{\mathcal{T}}

\def\cV{\mathcal{V}}

\def\mf{\mathfrak}

\def\nn{\nonumber}

\numberwithin{equation}{section}

\begin{document}
\pagestyle{myheadings}
\markboth{\textsc{\small }}{%
  \textsc{\small Subtracted Geometry From Harrison
   Transformations}}
  \addtolength{\headsep}{4pt}

%% \begin{titlepage}

\begin{flushright}
\texttt{AEI-2012-031}
\end{flushright}

\begin{centering}

  \vspace{0cm}

\textbf{\Large{Subtracted Geometry From Harrison Transformations}}

 \vspace{0.8cm}

  {\large Amitabh Virmani}

  \vspace{0.5cm}

\begin{minipage}{.9\textwidth}\small  \begin{center}
Max-Planck-Institut f\"ur Gravitationsphysik (Albert-Einstein-Institut) \\
Am M\"uhlenberg 1,
D-14476 Golm, Germany \\
{\tt virmani@aei.mpg.de}
\\ $ \, $ \\

\end{center}
\end{minipage}

\end{centering}

%\vspace{1cm}

\begin{abstract}
%  \vspace{0.8cm}
We consider the rotating non-extremal black hole of N$=$2 D$=$4 STU
supergravity carrying three magnetic charges and one electric charge. We show
that its subtracted geometry is obtained by applying a specific SO(4,4)
Harrison transformation on the black hole. As previously noted, the
resulting subtracted geometry is a solution of the N$=$2 S$=$T$=$U supergravity.
\end{abstract}
%\vfill

%\noindent \mbox{}
%\raisebox{-3\baselineskip}{%
%  \parbox{\textwidth}{ \mbox{}\hrulefill\\[-4pt]}}

\thispagestyle{empty}
\tableofcontents

%\newpage

\setcounter{equation}{0}

\section{Introduction}

Over the years there has been slow but steady progress in our understanding of
relations between black holes and two dimensional conformal field theories. Several
universal properties of black holes have been found to be related to universal
properties of 2d CFTs. String theory has provided significant
insights in this quest. Arguably, one of the most spectacular successes of
string theory is the Sen-Strominger-Vafa counting of the microscopic configurations, and thereby providing a
statistical mechanical explanation of the entropy of certain supersymmetric and
near-supersymmetric black holes \cite{Sen:1995in, Strominger:1996sh}.
Since then, many different types of black
holes have been studied and the agreement between the Bekenstein-Hawking
entropy and the statistical mechanical entropy has been shown to hold
in a variety of cases.

These achievements, very impressive as they are, need to be contrasted with
the challenge of microscopically understanding general non-extremal black holes. The methods advocated in \cite{Sen:1995in,
Strominger:1996sh} cannot be directly applied to such general settings. More recently, considerable progress has been made in
addressing general extremal black holes. These developments go under the name of
the Kerr/CFT correspondence \cite{Guica:2008mu}; see \cite{Bredberg:2011hp} for
a concise review and see \cite{Compere:2012jk} for a more comprehensive review\footnote{In these reviews further references on these and related developments
can also be found.}. Once again, these developments rely on certain specific
structure of extremal black holes, and cannot be directly generalized to non-extremal settings.
In the case of the Kerr/CFT, existence of the decoupled
near-horizon geometry is crucial. In settings far away from extremality one
cannot decouple the near-horizon region from the asymptotic region. As a result, it remains unclear how the considerations of
Kerr/CFT are useful for describing general non-extremal settings.

It comes as a surprise that even for black holes far away from extremality certain tantalizing
clues have been found for the presence of a conformal symmetry. It was observed in \cite{Castro:2010fd} that in certain low-energy
near-horizon regimes the dynamics of a probe scalar field enjoys
 a local hidden non-geometric SL(2, $\mathbb{R}$) $ \times $ SL(2, $\mathbb{R}$) symmetry. The precise meaning of this symmetry is a topic of future research, but the picture put forward in  \cite{Castro:2010fd} shows remarkable coherence. These hidden symmetries only appear in a region close enough to the
horizon. It has been suggested \cite{Bertini:2011ga, Cvetic:2011hp, Cvetic:2011dn, Cvetic:2012tr} that one can consistently deform the geometry of
an asymptotically flat black hole so that these hidden symmetries appear manifestly in the
deformed geometries. These geometries are dubbed ``subtracted
geometries.'' The subtracted geometries are not asymptotically flat. They are
supported by additional matter fields. In this work we explore these geometries and their relation to the original black holes.

The main aim of this paper is to establish that subtracted geometries can
be obtained from the original black hole by applying  solution generating
transformations. For concreteness we consider the case of  four-charge rotating
non-extremal four-dimensional asymptotically flat black holes of N=2 STU
supergravity. Moreover we restrict ourselves to the black hole carrying three
magnetic and one electric charge. This is just a choice; we expect our
considerations to straightforwardly apply to other combinations of in total four electric and
magnetic charges.

The motivation for looking at the 4d solution carrying three magnetic charges
(and one electric charge) is manifold. Not only we can perform a study of its
subtracted geometry, but also we can use it to perform various other studies;
most notably in relation to a string theory realization of the Kerr/CFT correspondence and black rings. It
was shown in \cite{G26} that the spinning magnetic one-brane of five-dimensional
minimal supergravity admits a near-horizon limit that smoothly interpolates between a
self-dual supersymmetric null orbifold of AdS$_3$ $\times$ S$^2$ and the
near-horizon limit of the extremal Kerr black hole times a circle. It is of interest to generalize this observation to a multicharge configuration. We present such a generalization in appendix \ref{onebrane}.

As for the construction of the rotating four-charge black hole carrying
three magnetic and one electric charge, there are several ways in which one can
approach this problem. The first, and perhaps also the most direct, approach
that comes to mind is to use boosts and string dualities.
One quickly realizes that to add three independent magnetic charges the number
of boosts and dualities steps required is in fact quite large (approximately 20).
To perform all these steps coherently is a computational challenge\footnote{A construction along
these lines of the spinning magnetic one-brane in five-dimensional U(1)$^3$ supergravity with three
independent M5 charges was attempted in \cite{Tanabe:2008vz}. However,
the author did not completely succeed in achieving this goal. The expressions presented in
\cite{Tanabe:2008vz} do not solve the supergravity equations.}.

There are other somewhat less computationally intensive possibilities. For
example, a second possibility is to perform an electro-magnetic duality in four-dimensional N=2 STU
supergravity and convert the two-electric two-magnetic rotating solution as presented in \cite{Chong:2004na} to three-magnetic and one-electric one.
Finally, a third possibility is to use the powerful machinery of three-dimensional hidden symmetries of the STU model to generate this solution.
It is the third path that was used to construct the solution carrying two electric and two magnetic charges \cite{Chong:2004na}.
In our opinion the second and the third routes are of almost equal computational
complexity. Since the approach of three-dimensional hidden symmetries also allows us to
relate to its subtracted geometry rather directly, we follow the third route in this paper.

For the ease of readability of the paper almost all technicalities related to
the construction of the solution are presented in appendices. Appendix
\ref{app:setup} presents the set-ups we work with in considerable detail. Here we
also present an implementation of the SO(4,4) nonlinear sigma
model. The group SO(4,4) is relevant because it is the group of hidden
symmetries of the N=2 STU supergravity when the theory is dimensionally reduced on a Killing vector.
The rest of the paper is organized as follows. We first construct the spinning
M5-M5-M5 solution in section \ref{sec:fourcharge}. We present it as a
configuration in five-dimensional U(1)$^3$ supergravity. Then we show how to add
the fourth charge.
In section \ref{subtracted} we obtain its subtracted geometry by applying a series of solution generating transformations. Three-dimensional sigma model fields
for the M5-M5-M5 solution are presented in appendix \ref{3dfields}.
Three dimensional fields for the subtracted geometry are presented in appendix \ref{3dfieldssubtracted}. We conclude in section \ref{conclusions}.

\section{Four-Charge Black Hole}
\label{sec:fourcharge}
Although four charge black holes of ungauged four dimensional supergravity
theories are well studied in the literature \cite{Cvetic:1996kv,
Chong:2004na,Cvetic:2009jn, Cvetic:2011dn}, to the best of our
knowledge expressions for all fields when the black hole carries three independent
magnetic charges have not been explicitly presented anywhere. We fill this gap
in this section. For many purposes, e.g., in relation to black rings, or in
relation to (0,4) MSW/D1-D5-KKM CFT \cite{MSW}, such a presentation is useful.

\subsection{M5-M5-M5}
\label{sec:m5cube}
We consider the M-theory frame and describe the configuration as a solution of
five-dimensional  U(1)$^3$ supergravity. Upon reducing over the string
direction we obtain a rotating 4d black hole carrying three independent magnetic charges. For various reasons we prefer to present the
5d lift of the 4d solution.

\subsubsection*{The Theory}
We follow the conventions in which the U(1)$^3$ supergravity Lagrangian takes
the a manifestly triality-invariant form
\begin{equation}
\label{eqn:Lagrangian5dmt}
\mathcal{L}_5 = R_5 \star_5 \mathbf{1}  - \frac{1}{2} G_{IJ} \star_5 dh^I \wedge  dh^J
- \frac{1}{2}G_{IJ}\star_5 F^I_{[2]}
\wedge F^J_{[2]}
+ \frac{1}{6}
C_{IJK} F^I_{[2]} \wedge F^J_{[2]} \wedge A^K_{[1]}.
\end{equation}
The symbol $C_{IJK}$ is
pairwise symmetric in its indices with $C_{123} = 1$ and is zero
otherwise.
The metric $G_{IJ}$ on the scalar moduli space is diagonal with entries $G_{II}
= (h^I)^{-2}$, where these scalars satisfy the constraint $h^1 h^2 h^3 = 1$.
This constraint must be solved before computing variations of the action to
obtain EOMs for various fields.

We construct the M5-M5-M5 solution using the familiar coset model solution
generating techniques. We reduce the theory \eqref{eqn:Lagrangian5dmt} on
commuting Killing vectors to three dimensions. We do this reduction first over
a spacelike Killing vector and then over a timelike Killing vector. The theory
reduces to 3d gravity coupled to SO(4,4)/(SO(2,2)$\times$SO(2,2)) non-linear sigma model.
Acting with an appropriate group elements of SO(4,4) on the Kerr string we get the non-extremal spinning magnetic one-brane of U(1)$^3$ supergravity.
Details on the set-up and the explicit form of the group element can be found in
appendix \ref{app:setup}. For five-dimensional  minimal supergravity such
constructions have been extensively discussed in our previous work \cite{G21,
G22, G23, G24, G25, G26}.

\subsubsection*{The Solution}

 Let $s_I = \sinh \alpha_I$ and $c_I = \cosh \alpha_I$ with $I = 1,2,3$, then the spinning magnetic one-brane with three-independent M5-charges is given as
\begin{equation}
\label{magnetic}
ds_5^2 = f^2 ( dz + {\check A_4}^0)^2 + f^{-1}( - e^{2U}(dt + \omega_3)^2 + e^{-2U}
ds_3^2(\mathcal{B})),
\end{equation}
where
\begin{equation}
\label{eqn:basemetric}
ds_3^2(\mathcal{B})  = \frac{\Delta_2}{\Delta} dr^2 + \Delta_2 d\theta^2 + \Delta \sin^2\theta d\phi^2,
\end{equation}
is the three-dimensional base metric obtained by reducing the Kerr string on $\partial_z$ first and then over $\partial_t$,  and
\begin{eqnarray}
\Delta\  & = & r^2-2m r + a^2, \qquad \Delta_2 = \Delta - a^2 \sin^2\theta \label{Delta2} \\
f^2 &=& 4\xi(\Omega_1 \Omega_2 \Omega_3)^{-2/3},  \qquad e^{4U} = \frac{{\Delta_2}^2}{\xi} \\
\omega_3 &= &c_1 c_2 c_3 \frac{2 a m r \sin^2\theta}{\Delta_2} d\phi,
\end{eqnarray}
are the metric functions appearing in the line element.  The rest of the metric functions take the form
\begin{eqnarray}
\xi &=& (r^2 + a^2 \cos^2\theta)^2 + 2m r(r^2+a^2\cos^2\theta)\left(\sum_{I=1}^{3} s_I^2\right) \nn \\
 & &  + 4 m^2r^2 \left(s_1^2 s_2^2 + s_2^2 s_3^2 + s_1^2 s_3^2 \right) + 4m^2(2
 m r - a^2 \cos^2\theta) \left(\prod_{I = 1}^{3}s_I^2\right),  \label{xi}
 \\
%\end{eqnarray}
%and
%\bea
{\Omega_1}  &=& 2(a^2\cos^2\theta +(r+2m s_2^2)(r+2m s_3^2)),\label{omega1} %\\
%{\Omega_2}  &=& 2(a^2\cos^2\theta +(r+2m s_1^2)(r+2m s_3^2)),\label{omega2}\\
%{\Omega_3}  &=& 2(a^2\cos^2\theta +(r+2m s_1^2)(r+2m s_2^2))\label{omega3}.
\eea
and cyclic permutations.
Furthermore we have
\begin{equation}
{\check A_4}^0 = \zeta^0(dt + \omega_3) + 2 s_1 s_2 s_3 \frac{am (r-2m)}{\Delta_2} \sin^2\theta d\phi,
\end{equation}
with
\begin{equation}
\zeta^0 = 4c_1c_2c_3s_1s_2s_3\frac{a^2 m^2 \cos^2\theta}{\xi}.
\end{equation}
The Maxwell potentials $A^I$'s of the five-dimensional theory take the form
\begin{equation}
A^I = \chi^I(dz + \check A_4^0) + \zeta^I (dt + \omega_3) + 2m s_I c_I
\frac{\Delta}{\Delta_2} \cos\theta d\phi
\label{vectors}
\end{equation}
with
\bea
\chi^1 &=& 4 c_1 s_2 s_3 \frac{a m \cos \theta}{\Omega_1} ,  \\
\zeta^1  &=& -2s_1 c_2 c_3 (r^2+a^2 \cos^2\theta +2rm s_1^2))  \frac{a m
\cos{\theta}}{\xi},
%\quad
%\chi^2 = 4 c_2 s_1 s_3 \frac{a m \cos \theta}{\Omega_2}, \quad \chi^3 = 4 c_3
% s_1 s_2 \frac{a m \cos \theta}{\Omega_3},
\eea
and obvious cyclic permutations.
Finally, the three scalars in the U(1)$^3$ theory take the form
\begin{equation}
h^I = (\Omega_1 \Omega_2 \Omega_3)^{1/3} \Omega_I^{-1}.
\end{equation}
The solution is sufficiently complicated, and it is non-trivial to
check that all supergravity equations are solved. We have checked that they are
solved \footnote{Our $\epsilon$ conventions are $\epsilon_{rx\phi z t} = + \sqrt{- \mathrm{det} \, g}$, with $x = \cos \theta$.}.

Setting any two of the three charges to zero, while keeping the angular momentum
non-zero, the resulting solution can be compared to reference
\cite{Harmark:1999xt}. In this special case the solution also admits a lift to
vacuum gravity in six dimensions. By setting the three charges equal the
solution can be compared with \cite{G21}. Certain physical properties of the
solution and its near horizon geometry in the extremal limit are studied in appendix \ref{onebrane}.

%\bea
%\zeta^2  &=& -2s_2 c_1 c_3 (r^2+a^2 \cos^2\theta +2rm s_2^2))  \frac{a m
% \cos{\theta}}{\xi},\\
%\zeta^3  &=& -2s_3 c_1 c_2 (r^2+a^2 \cos^2\theta +2rm s_3^2))  \frac{a m
% \cos{\theta}}{\xi}.
%\eea

\subsection{Adding the Fourth Charge}
By boosting the string configuration (\ref{magnetic}) in $(t,z)$, and then
dimensional reducing over the $z$ direction we obtain a four-charge four-dimensional black hole. The 4d black hole
carries three-magnetic charges and one-electric charge. From the
hidden symmetries point of view this procedure is equivalent to performing
\be
\cM_\rom{3-charge} \rightarrow \cM_\rom{4-charge} = g_4^\sharp \cdot
\cM_\rom{3-charge}  \cdot g_4, \label{fourcharge1}
\ee
with
\be
g_4 = \exp \left[ -\alpha_0 (E_{q_0} + F_{q_0}) \right]\label{fourcharge}.
\ee Here $\cM_\rom{3-charge}$
denotes the SO(4,4) coset matrix for the above three-charge
configuration\footnote{The notation $g^\sharp$ denotes a generalized
transposition. The transposition is defined on the generators of the $\mf{so}$(4,4) Lie algebra
by $\sharp(x) = - \tau(x) \: \forall \: x \in$ $\mf{so}$(4,4), where
$\tau$ is the involution of the Lie algebra that defines the coset. More details can be found in  appendix \ref{app:setup}.}. The explicit expressions for the resulting fields are fairly
complicated. For the case of two-electric and two-magnetic charges these expression are presented in full detail in \cite{Chong:2004na}. Fortunately, we will not need the explicit
expressions in what follows.

\section{Subtracted Geometry From Harrison Transformations}
\label{subtracted}
To obtain the subtracted geometry of the above described four-charge black hole
we act  on it with a series of solution generating transformations.  These
transformations perform the required Harrison boosts that give the
subtracted geometry. The precise sequence of transformations is somewhat involved. We perform them in a certain
specific order explained below to maintain the complexity of intermediate
expressions under control.

This investigation was  systematically initiated in \cite{Bertini:2011ga, Cvetic:2012tr}. In \cite{Cvetic:2012tr}
it was suggested that the subtracted geometry of the four-charge black hole can
be obtained by certain Harrison boosts. The subtracted geometries
of the Schwarzschild and Kerr solutions were obtained in
Einstein-Maxwell-Dilaton theories by applying certain infinite Harrison boosts.
The key technical observation we take from that work  is their equation (33),
i.e., that the Harrison boosts used are of the \emph{lower triangular} form.
From the point of view of the SO(4,4) Lie algebra this suggests that the specific Harrison
transformation that leads to the subtracted geometry of the four-charge black
hole belongs to `lowering' generators, more precisely to generators
corresponding to negative root vectors. This is indeed the case, as we explain
next.

\subsection{Charging, Harrison Boosts, and Scaling}
The most important transformation on the four-charge black hole to obtain its
subtracted geometry is of the form
\be
\cM_\rom{4-charge} \rightarrow \cM' = g_H^\sharp \cdot \cM_\rom{4-charge}
\cdot g_H,
\ee
with the Harrison transformation $g_H$
\be
g_H = \exp[(F_{p^1} + F_{p^2} + F_{p^3})]. \label{harrison}
\ee
Note that despite the fact that the four-charge black hole carries three
independent M5 charges, the Harrison boosts in \eqref{harrison} are by the same
`amount' in the $p^1,~p^2$ and $p^3$ `directions.' In all these three directions
the boosts are infinite, in the sense that the lowering generators
$F_{p^1},~F_{p^2}$ and $F_{p^3}$ are exponentiated with unit coefficients, in
line with \cite{Cvetic:2012tr}. Furthermore, note that we \emph{do not} apply a Harrison boost in the $q_0$ `direction.' This is
reminiscent of the near-extreme multi-charge black holes in the so-called dilute
gas approximation \cite{Cvetic:1997uw, Maldacena:1996ix}.

However, it so happens that performing the transformation \eqref{harrison} on the
four-charge black hole resulting from \eqref{fourcharge1}--\eqref{fourcharge} is quite
intricate to implement. To bypass this purely technical complexity we make
the following crucial observation: the generator that adds the fourth charge,
namely, $(E_{q_0} + F_{q_0})$ commutes with all three generators of the Harisson
boosts we want to perform $F_{p^1},~F_{p^2}$ and $F_{p^3}$. As a result the
transformation
\bea
\cM' &=& g_H^\sharp \cdot \cM_\rom{4-charge}
\cdot g_H \\  &=& g_H^\sharp \cdot g_4^\sharp \cdot
\cM_\rom{3-charge}  \cdot g_4 \label{harrison41}
\cdot g_H
\eea
is the same as doing
\be
\cM' = g_4^\sharp \cdot g_H^\sharp \cdot
\cM_\rom{3-charge}  \cdot g_H \label{harrison42}
\cdot g_4,
\ee
where we have commuted $g_4$ past $g_H$. Physically there is absolutely no
difference between  \eqref{harrison41} and \eqref{harrison42}, but computationally performing \eqref{harrison42} is
significantly simpler (at least in the way we have organized our computer
implementation of the SO(4,4) coset model).

This is not the end of the story. One also needs to perform a
further scaling transformation to get the subtracted geometry in precisely
the form given in \cite{Cvetic:2012tr}. This last transformation is as follows
\be
\cM_\rom{subtracted} = g_S^\sharp \cdot \cM'
\cdot g_S, \qquad g_S = \exp[-c_0 H_0 + c_1 H_1 + c_2 H_2 + c_3 H_3],
\label{scaling}
\ee
where $c_0,c_1,c_2,c_3$ are given below.
Having done all these solution generating transformations we need to change
variables along the line as suggested in \cite{Cvetic:2012tr} and choose the
parameters $c_0,c_1,c_2,c_3$ in \eqref{scaling} in a specific way. The choice
\bea
\alpha_I &=& -\frac{1}{2} \ln \left[\left( \Pi_c^2
-\Pi_s^2\right) d_I \right],
\label{choices1}  \\ 
\alpha_0 &=& \sinh^{-1}\left(\frac{\Pi_s}{\sqrt{\Pi_c^2 -\Pi_s^2}}\right),
\label{choices2} \\
c_0 &=& - \ln \left(\Pi_c^2 -\Pi_s^2 \right)- \frac{1}{4} \ln \left( d_1
d_2 d_3 \right), \label{choices3} \\
c_1 &=& \frac{1}{4} \ln \left[\frac{d_2 d_3}{d_1}\right], \quad c_2 \ \  = \ \
\frac{1}{4} \ln \left[\frac{d_1 d_3}{d_2}\right], \quad c_3 \ \  =
 \ \ \frac{1}{4} \ln \left[\frac{d_1 d_2}{d_3}\right],
\label{choices4} %\quad \mbox{and obvious cyclic permutations}
\eea
leads to the subtracted geometry of the four-charge black hole\footnote{The
explicit product expressions \cite{Cvetic:2011dn, Cvetic:2012tr}
$\Pi_c = \prod_{I=0}^{4} \cosh \alpha_I, \Pi_s = \prod_{I=0}^{4} \sinh
\alpha_I,
$ are not needed in our computations, because the final geometry is
parameterized solely in terms of $\Pi_c$ and $\Pi_s$. See also
footnote 3 of \cite{Cvetic:2012tr}.} as presented in \cite{Cvetic:2012tr}. The
three parameters $d_1, d_2, d_3$ introduced by the redefinitions
\eqref{choices1}--\eqref{choices4} are redundant from the point of view of the
final spacetime configuration. This happens in the following way. These
constants do appear in the sigma model fields, but  as `axionic shifts' of the dual potentials $\tilde \zeta_1,
\tilde \zeta_2, \tilde \zeta_3$ defined via equation \eqref{dual1}. These
constants also appear in the sigma model field $\sigma$ but in a very
special way so that $d \omega_3$ defined via equation \eqref{dual2} does not depend on them.
The final three dimensional sigma model fields are all given in appendix \ref{3dfieldssubtracted}. When the  spacetime configuration 
is constructed by dualizing these fields, parameters $d_1, d_2$, and $d_3$
completely disappear. We find these cancellations truly remarkable. As a
consequence of these cancellations we find the geometry precisely  as presented in
\cite{Cvetic:2012tr}. We discuss this geometry further in the next section. It 
is a solution of the N=2 D=4 S=T=U supergravity.
Perhaps a more general notion of
subtracted geometry is possible that is not contained in the S=T=U truncation.
Such a geometry would perhaps be a more natural candidate for a CFT dual that describes
black holes with all different electric and magnetic charges. This issue needs
further investigation.

We summarize. To obtain subtracted geometry of the four-charge black hole as
presented in  \cite{Cvetic:2012tr} we perform the following transformation on
the 3-charge black hole
\be
\cM_\rom{subtracted} = g_S^\sharp \cdot
g_4^\sharp \cdot g_H^\sharp \cdot
\cM_\rom{3-charge}  \cdot g_H \label{harrison43}
\cdot g_4 \cdot g_S.
\ee
For convenience and completeness all the resulting three-dimensional fields are
listed in appendix \ref{3dfieldssubtracted}. In the the next section we present
the final geometry in the four-dimensional language and compare it with the analysis of Cvetic and Gibbons.

\subsection{Resulting Geometry}
The resulting geometry in the four-dimensional language is most conveniently
expressed as
\begin{equation}
ds_4^2 = - e^{2U}(dt + \omega_3)^2 + e^{-2U}
ds_3^2(\mathcal{B}),
\end{equation}
where
\begin{equation}
\label{eqn:basemetric}
ds_3^2(\mathcal{B})  = \frac{\Delta_2}{\Delta} dr^2 + \Delta_2 d\theta^2 + \Delta \sin^2\theta d\phi^2,
\end{equation}
is the three-dimensional base metric obtained by reducing the Kerr black hole over $\partial_t$,  and
\be
\Delta   = r^2-2m r + a^2, \qquad \Delta_2 = \Delta - a^2 \sin^2\theta \label{Delta2n}.
\ee
Rewriting the four-dimensional metric in the form as in \cite{Cvetic:2011dn,
Cvetic:2012tr} we get
\be
ds^2_4 = - \left(\frac{1}{e^{-2U}\Delta_2}\right) \Delta_2 (dt + \omega_3)^2 + e^{-2U} \Delta_2 \left(\frac{dr^2}{\Delta} + d\theta^2 + \frac{\Delta}{\Delta_2}\sin^2 \theta d\phi^2\right).
\ee
The square of the factor $e^{-2U} \Delta_2$ is called the
subtracted conformal factor in  \cite{Cvetic:2011dn, Cvetic:2012tr}. From appendix \ref{3dfieldssubtracted} we read the value of $e^{-2U} \Delta_2$ to be
\be
e^{-2U} \Delta_2 = 2 m \sqrt{\tilde \Delta_s},
\ee
with
\be
\tilde \Delta_s = 2m r (\Pi_c^2 -\Pi_s^2) + 4 m^2 \Pi_s^2 - a^2 \cos^2 \theta (\Pi_c -\Pi_s)^2.
\ee
Our $4 m^2\tilde \Delta_s$ precisely corresponds to the subtracted conformal factor used in
\cite{Cvetic:2011dn, Cvetic:2012tr}. One form $\omega_3$ takes the form
\be
\omega_3 = \frac{2 m a (r(\Pi_c -\Pi_s) + 2 m \Pi_s)\sin^2\theta}{r^2 + a^2 \cos^2 \theta - 2 m r}d\phi.
\ee
For the four-dimensional axion and dilaton fields we find
\be
\chi^1 = \chi^2 = \chi^3 =  \frac{a(\Pi_c-\Pi_s) \cos \theta}{2m},
\ee
and
\be
y_1 =  y_2 =  y_3 = \frac{\sqrt{\tilde \Delta_s}}{2m},
\ee
which again precisely match with the expressions reported in
\cite{Cvetic:2012tr}, once we make a translation of conventions. Finally, the four dimensional vector fields take the form
\be
\check A^0_{[1]} =  \frac{4a m^2 \sin^2\theta (\Pi_c -\Pi_s)}{\tilde \Delta_s}
d\phi+ \frac{a^2 \cos^2\theta (\Pi_c -\Pi_s)^2 + 4 m^2 \Pi_c \Pi_s}{(\Pi_c^2 -
\Pi_s^2) \tilde \Delta_s} dt,
\ee
and
\be
\check A^1_{[1]} =\check A^2_{[1]} =\check A^3_{[1]},
\ee
\bea
\check A^1_{[1]} &=& 2 m \cos\theta \frac{2 m (2 m \Pi_s^2 + r (\Pi_c^2 -
\Pi_s^2)) - a^2 (\Pi_c - \Pi_s)^2}{\tilde \Delta_s} d\phi \nn \\
& & - \frac{a \cos \theta (2 m \Pi_s + r (\Pi_c -\Pi_s)}{\tilde \Delta_s} dt.
\eea
As far as the expressions for the vector fields can be compared with the
corresponding expressions in \cite{Cvetic:2012tr}, they perfectly match. Since
our vector field $\check A^1_{[1]}$ is magnetically sourced, whereas in \cite{Cvetic:2012tr} the
corresponding vector is electrically sourced a direct comparison is not
possible. We have explicitly checked that our subtracted solution solves all supergravity
equations. Furthermore, since the dilatons are all obtained to be equal and so
are the axions and the three vectors $\check A^1_{[1]} =\check A^2_{[1]} =\check
A^3_{[1]}$, the resulting solution is in fact a solution of the N=2 S=T=U
supergravity. This fact has been previously noted as well \cite{Cvetic:2011dn,
Cvetic:2012tr}.

\section{Conclusions}
\label{conclusions}
The key result of this paper is to show that the multicharge subtracted geometry
can be obtained via a series of solution generating transformations on the
original black hole field configuration. There are number of ways in which our
study can be extended. In this work we have concentrated on a four-charge
four-dimensional black hole carrying three magnetic charges and one electric charge. It
is fairly clear from our work how to implement the same procedure for the  black hole
carrying two electric and two magnetic charges. It can be a useful exercise
to fill in all details. In this regard understanding the precise
meaning of equations \eqref{choices1}--\eqref{choices4} is an important future direction. In the
same line of thought, it is interesting to explore a similar series of
transformations for the non-extremal rotating three-charge five-dimensional asymptotically flat black holes.

As explained in the previous work of Cvetic and Larsen \cite{Cvetic:2011hp, Cvetic:2011dn} and Cvetic and Gibbons \cite{Cvetic:2012tr},
expressions for the entropy and thermodynamic properties of the black hole are
preserved by the transformations leading to  subtracted geometries. It is hoped that the
dual CFT description  of the black hole is also somehow preserved. With
these motivations it is of interest to further study these geometries and in
particular to explore the existence of asymptotic Virasoro algebras in the
subtracted geometries. It will then be of interest to know how the asymptotic Virasoro symmetries get
transformed under the inverse solution generating transformations.
Such a line of investigation can teach us some general and important lessons about non-extremal
rotating black holes in string theory and their relation to two-dimensional
conformal field theories. We hope to report on some of these issues in our
future work.

\subsection*{Acknowledgements}
We thank Makoto Tanabe for useful correspondence regarding his work
\cite{Tanabe:2008vz}. We also thank Geoffrey Compere and Maria J Rodriguez for
discussions, and especially Josef Lindman H\"ornlund for several useful discussions
regarding Mathematica implementation of the SO(4,4) coset model. We are also
grateful to Geoffrey Compere and Teake Nutma for reading the manuscript and
making pertinent comments. After the first version of the paper was posted on
the ArXiv, we received useful
comments and suggestions from Guillaume Bossard, Finn Larsen, Oscar Varela, Paul
P.~Cook, and an anonymous referee, which have led to a
significant improvement in the results of section \ref{subtracted}.
\appendix

\section{The Set-Up}
\label{app:setup}
In this section we present the set-ups we work with. We also present certain
details on the implementation of the SO(4,4) coset model.
\subsection{A Chain of Dimensional Reductions}
Various relations
through dimensional reduction, truncations, and oxidations are presented.
All results of this section are already well known in the literature.
For this reason we shall be brief. The main purpose of this section is to set
the notation and conventions for the main text of the paper.
 \subsection*{Truncation of
IIB Theory on T$^4$}
\label{sec:IIB}
A well known consistent truncation of the IIB theory on a four-torus is as
follows
\bea
ds^2_{10, \, \rom{string}} = ds^2_6 + e^{\frac{\Phi}{\sqrt{2}}}
ds_4^2, \qquad
\Phi_{10} = \frac{\Phi}{\sqrt{2}}, \qquad  C^{\rom{RR}}_{[2]} = C_{[2]},
\eea
where $ds^2_4$ denotes the metric on the four-torus and $C^{\rom{RR}}_{[2]}$ is
the Ramond-Ramond two-form of the IIB theory. The rest of the IIB fields are set
to zero.
The two-form $C_{[2]}$ is the descendant from the
IIB Ramond-Ramond $C^{\rom{RR}}_{[2]}$ to six-dimensions. The resulting six-dimensional
theory contains a graviton, an antisymmetric tensor and a dilaton.
The bosonic part of the Lagrangian is \cite{Duff}
\be
{\cal L}_{6B} = R_6 \star_6 1 - \frac{1}{2}\star_6 d\Phi \wedge d  \Phi -
\frac{1}{2} e^{\sqrt{2} \Phi} \star_6 F_{[3]} \wedge F_{[3]},
\label{6d}
\ee
with the three-form field strength
$F_{[3]} =
dC_{[2]}$. Upon further dimensional
reduction on a two-torus the six-dimensional theory (\ref{6d}) reduces to the
N=2 STU model in four-dimensions. We present certain details of this
construction in the following.

\subsection*{Five-dimensional U(1)$^3$ supergravity}
\label{sec:U1cubic}
M-theory on six-torus admits a truncation to five-dimensional U(1)$^3$
supergravity. For relevant details see e.g. \cite{Emparan:2006mm}.
It can also be obtained by circle reduction of the Lagrangian \eqref{6d}.
We follow this route here. Using the standard Kaluza-Klein ansatz for the
six-dimensional fields \cite{Pope}
\bea
\label{metric6d}
ds^2_6 &=& e^{-\sqrt{\frac{3}{2}} \Psi} (dz_6 + A_{[1]}^1)^2 +
e^{\frac{1}{\sqrt{6}} \Psi} ds^2_5 \\
F_{[3]} &=& F_{[3]}^{\rom{(5d)}} + dA_{[1]}^2\wedge (dz + A_{[1]}^1)
\eea
with
\be
  F_{[3]}^{\rom{(5d)}} = dC_{[2]}^{\rom{(5d)}} -dA^2_{[1]}\wedge A^1_{[1]},
\ee
we obtain the following five-dimensional Lagrangian
\bea
{\cal L}_5 &=& R_5 \star_5 1 - \frac{1}{2} \star_5 d \Phi \wedge d \Phi -
\frac{1}{2} \star_5 d \Psi \wedge d \Psi - \frac{1}{2} e^{-2
\sqrt{\frac{2}{3}} \Psi} \star_5 F^1_{[2]}\wedge F^1_{[2]}
\nonumber
\\
& & - \frac{1}{2} e^{- \sqrt{\frac{2}{3}}\Psi + \sqrt{2} \Phi}
\star_5 F^{\rom{(5d)}}_{[3]} \wedge  F^{\rom{(5d)}}_{[3]} - \frac{1}{2}
e^{\sqrt{\frac{2}{3}}\Psi + \sqrt{2} \Phi} \star_5 F^2_{[2]} \wedge
F^2_{[2]},
\label{5dLag}
\eea
where $F^I_{[2]} = dA^I_{[1]}$ and $I=1,2$. Now, in five-dimensions the two-form
$C^{\rom{(5d)}}_{[2]}$ is dual to a one-form, which we denote as $A^3_{[1]}$.
After this dualization we end up with three one-forms in five-dimensions. We
 use the notation $A_{[1]}^I$, where now the index $I$ runs as $I=1,2,3$. We
 see the triality structure of the U(1)$^3$ supergravity emerging. The Chern-Simons term of the
U(1)$^3$ supergravity is also obtained through this dualization.

To see this,
recall that in the process of dualisation, Bianchi identities exchange
role with the equations of motion. The Bianchi identity for
$F^{\rom{(5d)}}_{[3]}$ is
\be
d F^{\rom{(5d)}}_{[3]} + F^2_{[2]} \wedge F^1_{[2]} = 0. \label{BI3}
\ee
The easiest way to do the dualization is to introduce $A^3_{[1]}$ as a Lagrange
multiplier for the Bianchi identity \eqref{BI3}.
Adding the appropriate  Lagrange multiplier term
to \eqref{5dLag} we get
\be
{\cal L}_{5}'  = {\cal L}_5 + A_{[1]}^3 \wedge (d F^{\rom{(5d)}}_{[3]} +
F^2_{[2]}\wedge F^1_{[2]}). \label{5dLagPrime}
\ee
As the next step, we treat the field strength $F^{\rom{(5d)}}_{[3]}$ as a
fundamental fields.
Varying ${\cal L}_{5}'$ with respect to  $F^{\rom{(5d)}}_{[3]}$ we find
\be
F^3_{[2]} - e^{-\sqrt{\frac{2}{3}} \Psi + \sqrt{2} \Phi } \star_5
F^{\rom{(5d)}}_{[3]} = 0.
\ee
Substituting this back into the Lagrangian \eqref{5dLagPrime}, we get
\bea
{\cal L}_5' &=& R_5 \star_5 1 - \frac{1}{2} \star_5  d \Phi \wedge  d \Phi -
\frac{1}{2} \star_5 d \Psi \wedge d \Psi \nonumber \\
& & - \frac{1}{2} e^{-2 \sqrt{\frac{2}{3}} \Psi} \star_5 F^1_{[2]} \wedge
  F^1_{[2]} - \frac{1}{2} e^{\sqrt{\frac{2}{3}}\Psi + \sqrt{2} \Phi}  \star_5
  F^2_{[2]} \wedge F^2_{[2]} \nonumber \\
&&  - \frac{1}{2} e^{\sqrt{\frac{2}{3}}\Psi - \sqrt{2} \Phi} \star_5
F^3_{[2]}\wedge F^3_{[2]} + A^3_{[1]} \wedge F^2_{[2]} \wedge F^1_{[2]}.
\label{Lag5dU13}
\eea

Lagrangian \eqref{Lag5dU13} is equivalent to five-dimensional U(1)$^3$
supergravity with the parameterization of the real special manifold as
\be
h^1 = e^{\sqrt{\frac{2}{3}} \Psi}, \quad h^2 = e^{-\sqrt{\frac{1}{6}} \Psi -\sqrt{\frac{1}{2}} \Phi}, \quad h^3 = e^{-\sqrt{\frac{1}{6}} \Psi + \sqrt{\frac{1}{2}} \Phi}.
\ee Clearly $h^1 h^2 h^3 = 1$.   A manifestly
triality-invariant form now be written as (we drop the prime on $\mathcal{L}_5'$ from now on)
\begin{equation}
\label{eqn:Lagrangian5d}
\mathcal{L}_5 = R_5 \star_5 \mathbf{1}  - \frac{1}{2} G_{IJ} \star_5 dh^I \wedge  dh^J
- \frac{1}{2}G_{IJ}\star_5 F^I_{[2]}
\wedge F^J_{[2]}
+ \frac{1}{6}
C_{IJK} F^I_{[2]} \wedge F^J_{[2]} \wedge A^K_{[1]}.
\end{equation}
The symbol $C_{IJK}$ is
pairwise symmetric in its indices with $C_{123} = 1$ and is zero
otherwise.
The metric $G_{IJ}$ on the scalar moduli space is diagonal with entries $G_{II}
= (h^I)^{-2}.$

For completeness, let us also write the six-dimensional field strength $F_{[3]}$
in terms of the five-dimensional fields introduced above. We obtain
\be
F_{[3]}  = - (h^3)^{-2} \star_5 dA^3_{[1]} + dA^2_{[1]} \wedge (dz_6 +
A^1_{[1]}).
\label{RR3}
\ee
Together with (\ref{metric6d}), equation \eqref{RR3} allows us to uplift any
solution of five-dimensional U(1)$^3$ supergravity to the IIB
theory. Examining the RR 3-form \eqref{RR3} reveals that the electric charge
that couples to the two-form $F^3_{[2]}$ arises from D5-branes wrapped on $T^5$:
$(z_6, z_7, z_8, z_9, z_{10})$. Similarly, the electric charge that couples to
the two-form $F^2_{[2]}$ arises from D1-branes wrapped along the $z_6$-circle.
The appearance of $A^1_{[1]}$ in the metric reveals that electric charge that
couples to $F^1_{[2]}$ arises from momentum (P) around the $z_6$-circle. The
interpretation of magnetic couplings is readily
obtained.
The M-theory interpretation of these couplings is reviewed at several places. See e.g.
\cite{Emparan:2006mm}.

\subsection*{Four-dimensional STU Model}

Further dimensional reduction of the five-dimensional U(1)$^3$ supergravity to
four dimensions gives the so-called STU model. The STU model is a
particular N=2 supergravity in four dimensions coupled to three vector
multiplets.

To fix our notation we quickly review here the N=2
supergravity action. Four-dimensional N=2 supergravity coupled to $n_v$
vector-multiplets is governed by a prepotential function $F$ depending on $(n_v + 1)$ complex
scalars $X^\Lambda$ $(\Lambda=0,1, \ldots,\,n_v)$. The bosonic degrees of freedom are the metric $g_{\mu \nu}$, the
complex scalars $X^\Lambda$ and a set of $(n_v+1)$ one-forms $\check
A^\Lambda_{[1]}$.
The bosonic part of the action is given as \cite{Ceresole:1995jg}
\be
\mathcal{L}_4 =  R \star_4 \mathbf{1} - 2 g_{I\bar{J}} \star_4 dX^I \wedge
d\bar{X}^{\bar{J}} + \frac{1}{2} \check F^{\Lambda}_{[2]} \wedge \check
G_{\Lambda}{}_{[2]},
\label{N=2}
\ee
where $\check F^\Lambda_{[2]} = d \check A^\Lambda_{[1]}$. The ranges of the
indices are $I,J = 1, \ldots, n_v$, and  $ g_{I \bar{J}} = \partial_I
\partial_{\bar{J}} K$ is the K\"ahler metric with the K\"ahler potential $K = -
\log \left[ -i (\bar{X}^\Lambda F_\Lambda - \bar{F}_\Lambda X^\Lambda)
\right].$ The two-forms $\check G_\Lambda{}_{[2]}$ are defined as \be \check
G_\Lambda{}_{[2]} = (\mbox{Re} N)_{\Lambda \Sigma} \check F^\Sigma_{[2]} +
(\mbox{Im} N)_{\Lambda \Sigma}\star_4 \check F^\Sigma_{[2]}~, \ee where the
complex symmetric matrix $N_{\Lambda \Sigma}$ is constructed from the
prepotential $F(X)$ as \be \label{matrixN} N_{\Lambda \Sigma} =
\bar{F}_{\Lambda \Sigma} + 2 i \frac{(\mbox{Im}F \cdot X)_\Lambda(\mbox{Im}F
\cdot X)_\Sigma}{X \cdot\mbox{Im}F \cdot X}~,\ee with $F_\Lambda =
\partial_\Lambda F$ and $F_{\Lambda \Sigma} = \partial_\Lambda \partial_\Sigma
F.$  For the system we are interested in $n_v = 3$ and the prepotential is
\be
F(X) = - \frac{X^1 X^2 X^3}{X^0}.\label{preSTU}
\ee

Let us now make contact of this Lagrangian with the circle reduction of the
five-dimensional U(1)$^3$ supergravity. We parametrize our
five-dimensional space-time as
\begin{equation}
ds^2_5 = f^{2}(dz + \check A^0_{[1]})^2 + f^{-1} ds^2_4,
\end{equation}
and the vectors as
\begin{equation}
A^I_{[1]} = \chi^I(dz+ \check A^0_{[1]})+ \check A^I_{[1]} .
\end{equation}
Together the graviphoton $\check A^0_{[1]}$ and the vectors $\check A^I_{[1]}$
form a symplectic vector $\check A^{\Lambda}_{[1]}$ with $\Lambda = 0,1,2,3$ in
four dimensions.

Upon circle reduction of the above 5d theory we obtain (with $\check
F_{[2]}^\Lambda = d \check A_{[1]}^\Lambda$)
\bea
\mathcal{L}_4 &=&  R \star_4 \mathbf{1} - \frac{1}{2}G_{IJ} \star_4 d h^I \wedge
d h^J  - \frac{3}{2f^2}\star_4 df \wedge df \nonumber  - \frac{f^3}{2}\star_4
\check F_{[2]}^0 \wedge \check F_{[2]}^0 \\
&& - \frac{1}{2 f^2}G_{IJ} \star_4 d \chi^I \wedge d \chi^J  - \frac{f}{2}
G_{IJ} \star_4 (\check F_{[2]}^I +\chi^I \check F_{[2]}^0) \wedge (\check
F_{[2]}^J +\chi^J \check F_{[2]}^0) \label{STUaction}
\\
&&+\frac{1}{2}C_{IJK}\chi^I \check F_{[2]}^J  \wedge \check F_{[2]}^K  +
\frac{1}{2} C_{IJK}\chi^I \chi^J \check F_{[2]}^0  \wedge \check F_{[2]}^K + \frac{1}{6}
C_{IJK} \chi^I \chi^J \chi^K \check F_{[2]}^0 \wedge \check F_{[2]}^0\, .
 \nn
\eea
The scalars $\chi^I$ and $h^I$ combine to form the complex scalars $z^I =
X^I/X^0$ in the STU theory according to
$z^I  = - \chi^I + i f h^I.$ Using the gauge fixing condition $X^0 = 1$
and the replacement $X^I \rightarrow z^I$ the action \eqref{N=2} for the
prepotential \eqref{preSTU} can be shown to be exactly equivalent
to the action \eqref{STUaction}. In order to perform the above
computation we found appendix A of reference \cite{Cardoso:2007rg} useful.

\subsection{SO(4,4)/(SO(2,2) $\times$ SO(2,2)) Coset Model in 3d}
In this section we discuss how to obtain the SO(4,4)/(SO(2,2) $\times$
SO(2,2)) coset model in three-dimensions by performing further dimensional reduction
over time direction of the STU action \eqref{STUaction}. We
 parametrize our four-dimensional space-time as
\begin{equation}
ds^2_4 =  -e^{2U}(dt+\omega_3)^2 + e^{-2U}ds_3^2,
\end{equation}
and the four-dimensional vectors as
\begin{equation}
\check A^\Lambda_{[1]} = \zeta^\Lambda(dt+\omega_3) + A_3^\Lambda,
\end{equation}
where $\omega_3$ and $A_3^\Lambda$ are one-forms in three-dimensions.

Following \cite{Gaiotto:2007ag, Bossard:2009we} we dualize the three dimensional
vectors as
\begin{equation}
-d\tilde{\zeta}_{\Lambda}  = e^{2U}(\mbox{Im} N)_{\Lambda \Sigma} \star_3 (d{A_3}^{\Sigma} + \zeta^{\Sigma} d\omega_3)+ (\mbox{Re} N)_{\Lambda \Sigma}d\zeta^{\Sigma}
\label{dual1}
\end{equation}
where $\tilde{\zeta}_{\Lambda}$ are pseudo-scalars. Similarly we define the
pseudo-scalar $\sigma $ dual to $\omega_3$ as
\begin{equation}
-d\sigma = 2 e^{4U} \star_3 d \omega_3 - \zeta^{\Lambda} d \tilde{\zeta}_{\Lambda} + \tilde{\zeta}^{\Lambda} d \zeta_{\Lambda} .
\label{dual2}
\end{equation}
 The full set of three-dimensional scalar fields are now $\varphi^a =
 \{U,z^I,\bar z^I,\zeta^\Lambda,\tilde \zeta_\Lambda,\sigma \}$. The resulting
 three-dimensional Lagrangian takes the form
\bea
\cL_3 = R \star_3 \mathbf{1} -\frac{1}{2} G_{ab} \partial \varphi^a \partial
\varphi^b,
\eea
 where the target space Lorentzian manifold parametrized by scalars
 $\varphi^a$ is of signature $(8,8)$. It is an analytic continuation of the
 c-map of Ferrara and Sabharwal
 \cite{Ferrara:1989ik}. The metric in our conventions
 is\footnote{Our conventions are identical to that of \cite{Bossard:2009we}.
 There is a minor typo of a factor of $1/2$ in equation (4.4) of
 \cite{Bossard:2009we}.}
\bea
 G_{ab}d\varphi^a d\varphi^b &=& 4 dU^2 + 4 g_{I \bar{J}}dz^I dz^{\bar J} +
 \frac{1}{4}e^{-4U} \left( d\sigma +  \tilde \zeta_\Lambda d \zeta^\Lambda -
 \zeta^\Lambda d \tilde \zeta_\Lambda \right)^2 \label{cmapst} \\
 && \hspace{-2cm} + e^{-2U}\left[ (\mbox{Im} N)_{\Lambda \Sigma}d\zeta^\Lambda
 d\zeta^\Sigma + ((\mbox{Im} N)^{-1})^{\Lambda \Sigma} \left( d\tilde
 \zeta_\Lambda +(\mbox{Re} N)_{\Lambda \Xi} d\zeta^\Xi \right)  \left( d\tilde
 \zeta_\Sigma +(\mbox{Re} N)_{\Sigma \Xi} d\zeta^\Xi \right) \nonumber \right].
\eea
This symmetric space can  be parametrized in the Iwasawa gauge by the coset
element \cite{Bossard:2009we}
\be
\label{iwa}
\cV = e^{- U \, H_0} \cdot \left(
\prod_{I=1,2,3}
e^{-\frac{1}{2} (\log y^I) H_I} \cdot e^{ - x^I E_I} \right) \cdot
e^{-\zeta^\Lambda E_{q_\Lambda}-  \tilde \zeta_\Lambda E_{p^\Lambda}}\cdot
e^{-\frac{1}{2}\sigma E_0},
 \ee
 where we use the notation $z^I = x^I + i y^I$ (so, $y^I = f h^I$, $x^I =
 -\chi^I$).
 The Iwasawa parameterization only covers an open subset of the full manifold. This is because the target space is
 not precisely the c-map but an analytic continuation of it.
The metric \eqref{cmapst} is
obtained from the Maurer-Cartan one-form $\theta= d\cV \cdot \cV^{-1}$,
\be
G_{ab} d\varphi^a d\varphi^b = \mathrm{Tr}( P_*\; P_* )\ ,\quad
P_* = \frac12 ( \theta + \eta'\, \theta^T {\eta'}^{-1})\ ,\quad
\eta' = {\rm diag}(-1,-1,1,1,-1,-1,1,1),
 \label{dsp}
\ee
where $\eta'$ is the quadratic form preserved by SO(2,2)$\times$SO(2,2). The
matrix $\cM$ is defined as
$\cM = (\cV^\sharp) \cV$,
with $\theta^\sharp = \eta' \theta^T {\eta'}^{-1}$ for all $\theta \in
\mf{so}(4,4)$. For convenience we
explicitly list the matrix representation of SO(4,4) in appendix
\ref{app:matrix}.
\subsection{Matrix representation of $\mf{so}$(4,4) Lie algebra}
\label{app:matrix}
An explicit realization of
the generators of $\mf{so(4,4)}$ is as follows. Calling $E_{ij}$ the $8 \times
8$ matrix with 1 in the $i$-th row and $j$-th column and 0 elsewhere, the
$\mf{so(4,4)}$ generators in the fundamental representation are given by
\bea
H_0 = E_{33} + E_{44} - E_{77} - E_{88} && H_1 = E_{33} - E_{44} - E_{77} + E_{88}\nn \\
H_2 = E_{11} + E_{22} - E_{55} - E_{66} && H_3 = E_{11} - E_{22} - E_{55} + E_{66}
\eea
\bea
E_0 = E_{47} - E_{38} && E_1 = E_{87} - E_{34}\nn \\
E_2 = E_{25} - E_{16} && E_3 = E_{65}-E_{12}  \\
F_0 = E_{74} - E_{83} &&  F_1 = E_{78} - E_{43}\nn \\
F_2 = E_{52} - E_{61} && F_3 = E_{56} - E_{21}\\
E_{q{}_0} = E_{41} - E_{58} && E_{q{}_1} = E_{57} - E_{31} \nn \\
E_{q{}_2} = E_{46} - E_{28} && E_{q{}_3} = E_{42} - E_{68}   \\
F_{q{}_0} = E_{14} - E_{85} && F_{q{}_1} = E_{75} - E_{13} \nn \\
F_{q{}_2} = E_{64} - E_{82} && F_{q{}_3} = E_{24} - E_{86}      \\
E_{p{}^0} = E_{17} - E_{35} && E_{p{}^1} = E_{18} - E_{45}\nn  \\
E_{p{}^2} = E_{67} - E_{32} && E_{p{}^3} = E_{27} - E_{36}      \\
F_{p{}^0} = E_{71} - E_{53} && F_{p{}^1} = E_{81} - E_{54} \nn  \\
F_{p{}^2} = E_{76} - E_{23} && F_{p{}^3} = E_{72} - E_{63}.
\eea
This basis of representation is identical to the one given in
\cite{Bossard:2009we}. For more details we refer the reader to this reference.  Other implementations of the SO(4,4) coset model can be found in \cite{Chong:2004na, Bossard:2011kz, Gal'tsov:2008nz}.

\subsection{Group element for the M5-M5-M5 black hole}
On the Kerr matrix $\cM_\rom{Kerr}$ we act with the group element
\be
g = \exp \left[  \alpha_1 (E_{p{}^1} + F_{p{}^1}) \right] \cdot \exp \left[  \alpha_2 (E_{p{}^2} + F_{p{}^2}) \right]
\cdot \exp \left[  \alpha_3 (E_{p{}^3} + F_{p{}^3}) \right],
\label{groupEle}
\ee
as
\be
\cM_\rom{Kerr} \rightarrow \cM_\rom{3-charge} = g^\sharp \cdot \cM_\rom{Kerr} \cdot g.
\ee
Reading off the new scalars from the new matrix $\cM_\rom{3-charge}$
and performing the inverse dualization through \eqref{dual1}--\eqref{dual2}
we obtain the spinning magentic one-brane of five-dimensional U(1)$^3$ supergravity as presented in section \ref{sec:fourcharge}.

\section{Three Dimensional Fields: 4d Asymptotically Flat}
\label{3dfields}
We list all the resulting three-dimensional fields obtained after the action of the group element \eqref{groupEle} on the coset matrix $\cM_\rom{Kerr}$:
\bea
x_1 &=& - 4 c_1 s_2 s_3 \frac{a m \cos \theta }{\Omega_1}, \\
x_2 &=& - 4 c_2 s_3 s_1 \frac{a m \cos \theta }{\Omega_2}, \\
x_3 &=& - 4 c_3 s_1 s_2 \frac{a m \cos \theta }{\Omega_3},
\eea
\bea
y_1 = \frac{2}{\Omega_1} \sqrt{\xi}, \qquad y_2 = \frac{2}{\Omega_2} \sqrt{\xi}, \qquad y_3 = \frac{2}{\Omega_3} \sqrt{\xi},
\eea
\bea
\zeta^0 &=& 4 c_1 c_2 c_3  s_1 s_2 s_3 \frac{a^2 m^2 \cos^2 \theta }{\xi}, \\
\zeta^1 &=& - 2 s_1 c_2 c_3 (r^2 + a^2 \cos^2 \theta + 2 m r s_1^2)\frac{a m \cos \theta }{\xi}, \\
\zeta^2 &=& - 2 s_2 c_3 c_1 (r^2 + a^2 \cos^2 \theta + 2 m r s_2^2)\frac{a m \cos \theta }{\xi},\\
\zeta^3 &=& - 2 s_3 c_1 c_2 (r^2 + a^2 \cos^2 \theta + 2 m r s_3^2)\frac{a m \cos \theta }{\xi},
\eea
\bea
\tilde \zeta_0 &=& 2 m a \cos \theta s_1 s_2 s_3  \frac{\Delta_2}{\xi}, \\
\tilde \zeta_1 &=&  \frac{m c_1 s_1}{\xi} \left(4 m a^2 \cos^2 \theta s_2^2 s_3^2 - r \Omega_1\right), \\
\tilde \zeta_2 &=&  \frac{m c_2 s_2}{\xi} \left(4 m a^2 \cos^2 \theta s_1^2 s_3^2 - r \Omega_2\right),\\
\tilde \zeta_3 &=&  \frac{m c_3 s_3}{\xi} \left(4 m a^2 \cos^2 \theta s_1^2 s_2^2 - r \Omega_3\right),
\eea
and finally
\bea
e^{2U} &=& \frac{\Delta_2}{\sqrt{\xi}}, \\
\sigma &=& - 4 m a \cos \theta c_1 c_2 c_3 \frac{r^2 + a^2 \cos^2 \theta + m r (s_1^2 + s_2^2 + s_3^2)}{\xi},
\eea
where $\Omega_1$,$\Omega_2$, and $\Omega_3$ are defined in
\eqref{omega1} and $\Delta_2$ and $\xi$ are defined respectively
in \eqref{Delta2} and \eqref{xi}. Finally, $c_1 = \cosh \alpha_1$, $c_2 = \cosh
\alpha_2$, $c_3 = \cosh \alpha_3$ and $s_1 = \sinh \alpha_1$, $s_2 = \sinh
\alpha_2$, $s_3 = \sinh \alpha_3$.

\section{Three Dimensional Fields: 4d Subtracted Geometry}
\label{3dfieldssubtracted}
Here we list all the resulting three-dimensional fields obtained after the action of the group element \eqref{harrison43}
with the choices \eqref{choices1}--\eqref{choices4}
on the coset matrix $\cM_\rom{3-charge}$ (here $x=\cos \theta$):
\be
x_1 = x_2 = x_3 = - \frac{ax (\Pi_c-\Pi_s)}{2m}.
\ee
Defining
\be
\tilde \Delta_s = 2m r (\Pi_c^2 -\Pi_s^2) + 4 m^2 \Pi_s^2 - a^2 x^2 (\Pi_c -\Pi_s)^2,
\ee
we have
\be
y_1 =  y_2 =  y_3 = \frac{\sqrt{\tilde \Delta_s}}{2m},
\ee
\be
\zeta^0 = \frac{4 m^2 \Pi_c \Pi_s + a^2 x^2 (\Pi_c -\Pi_s)^2}{\tilde \Delta_s (\Pi_c^2 -\Pi_s^2)},
\ee
\be
\zeta^1 = \zeta^2 = \zeta^3 =  - \frac{a x (2 m \Pi_s + r (\Pi_c - \Pi_s))}{\tilde \Delta_s},
\ee
\be
\tilde \zeta_0 = \frac{ax}{2m \tilde \Delta_s}
\left(
(\Pi_c-\Pi_s)^2(\Pi_c+\Pi_s) (r^2 + a^2 x^2)
-2 m r (\Pi_c^3 - 2 \Pi_c \Pi_s^2 + \Pi_s^3)
- 4 m^2 \Pi_c \Pi_s^2
\right),
\ee
\be
\tilde \zeta^1 \neq \tilde \zeta^2 \neq \tilde \zeta^3,
\ee
\bea
\tilde \zeta^1 &=&
\frac{1}{2 \tilde \Delta_s (\Pi_c^2 - \Pi_s^2)} \Big{[}-2 m r (\Pi_c^2 - \Pi_s^2) (1 + \Pi_c^2 - 3 \Pi_s^2) + 4 m^2 \Pi_s^2 ( \Pi_s^2 - \Pi_c^2 -1) \nn \\
 & & + (\Pi_c - \Pi_s)^2 (2 r^2 (\Pi_c + \Pi_s)^2 + a^2 x^2 (1 + (\Pi_c + \Pi_s)^2))\Big{]}
 + \frac{d_1-1}{2 d_1 \left(\Pi_c^2-\Pi_s^2\right)}\\
 \tilde \zeta^2 &=&
\frac{1}{2 \tilde \Delta_s (\Pi_c^2 - \Pi_s^2)} \Big{[}-2 m r (\Pi_c^2 - \Pi_s^2) (1 + \Pi_c^2 - 3 \Pi_s^2) + 4 m^2 \Pi_s^2 ( \Pi_s^2 - \Pi_c^2 -1) \nn \\
 & & + (\Pi_c - \Pi_s)^2 (2 r^2 (\Pi_c + \Pi_s)^2 + a^2 x^2 (1 + (\Pi_c + \Pi_s)^2))\Big{]}
 + \frac{d_2-1}{2 d_2 \left(\Pi_c^2-\Pi_s^2\right)}\\
 \tilde \zeta^3 &=&
\frac{1}{2 \tilde \Delta_s (\Pi_c^2 - \Pi_s^2)} \Big{[}-2 m r (\Pi_c^2 - \Pi_s^2) (1 + \Pi_c^2 - 3 \Pi_s^2) + 4 m^2 \Pi_s^2 ( \Pi_s^2 - \Pi_c^2 -1) \nn \\
 & & + (\Pi_c - \Pi_s)^2 (2 r^2 (\Pi_c + \Pi_s)^2 + a^2 x^2 (1 + (\Pi_c + \Pi_s)^2))\Big{]}
 + \frac{d_3-1}{2 d_3 \left(\Pi_c^2-\Pi_s^2\right)}
\eea
and finally
\bea
\s &=& \frac{a x}{2 m (\Pi_c^2 - \Pi_s^2) \tilde \Delta_s} \Big{[}
(r^2 + a^2 x^2) (\Pi_c - \Pi_s)^2 (\Pi_c + \Pi_s) -
 m r (\Pi_c - \Pi_s) (3 +
    3 \Pi_c^2 - \Pi_s^2) \nn \\
& & \qquad \qquad \qquad \qquad -
 2 m^2 \Pi_s (3 + \Pi_c^2 + \Pi_s^2)
\Big{]} \nn \\ 
&& + 
\left[\frac{(3 d_1 d_2 d_3-d_1 d_2 - d_1 d_3 - d_2 d_3)}{2
    d_1 d_2 d_3}  \right]
\frac{a x  
(2 m \Pi_s+r (\Pi_c-\Pi_s))}{
    (\Pi_c^2-\Pi_s^2) 
    %\left(-a^2 x^2 (\Pi_c-\Pi_s)^2+4 m^2 \Pi_s^2+2 m r
    %\left(\Pi_c^2-\Pi_s^2\right)\right)
    \tilde \Delta_s}
,
\eea
\bea
e^{2U} &=& \frac{r^2 + a^2 x^2 -2 m r}{2 m \sqrt{\tilde \Delta_s}}.
\eea

\section{Magnetic One Brane of U(1)$^3$ Theory}
\label{onebrane}
We provide an analysis of physical properties and near horizon limit of the
rotating magnetic string (\ref{magnetic}).
\subsection{Physical Properties}
 From the $g^{rr}$ component of the metric it
is seen that the solution has a regular outer horizon at $r= r_{+} :=  m + \sqrt{m^2 -a^2}$ and an inner horizon at $r= r_{-} := m - \sqrt{m^2 -a^2}$. The extremal limit is when the two horizons coincide, i.e., $m = a$.
The ADM stress tensor takes the form
\be
T_{tt} = \frac{m}{2G} \left( 2 +  s_1^2 + s_2^2 + s_3^2 \right), \quad  T_{zz} =
-\frac{m}{2G} \left( 1 + s_1^2 + s_2^2 + s_3^2 \right), \quad T_{tz} = 0,
\ee
where $T_{tt}$ and $T_{tz}$ are respectively the energy and linear momentum density along the string. $T_{zz}$ is the pressure density; the ADM tension is $\cT = - T_{zz}$.
Physical properties of the solution such as mass, inner and outer horizon areas, angular momentum, and angular velocities can be straightforwardly calculated. For the asymptotic quantities one finds
\bea
M &=&  2 \pi R  T_{tt} = \frac{\pi m R}{G} \left( 2 + s_1^2 + s_2^2 + s_3^2 \right), \label{mass_magnetic} \\
P_{z} &=& 2 \pi R T_{tz} = 0, \quad
J_{\phi} = \frac{2 \pi R m a}{G} c_1 c_2 c_3,
\eea
where $z \sim z + 2 \pi R$. For quantities at the outer ($r=r_+$) and inner ($r=r_-$) horizon one finds
\be
\Omega^\pm_{\phi} = \frac{a}{2 c_1 c_2 c_3 m r_\pm},   \quad
v^\pm_{z} = - \frac{a^2 s_1 s_2 s_3}{r_\pm^2 c_1 c_2 c_3}, \label{omegam}  \quad
A^\pm_{H} =  8 \pi^2 R \: (r_\pm^2 +a^2) c_1 c_2 c_3.
\ee
Temperatures of the inner and the outer horizons can be calculated from surface gravities,
\be
T^\pm_{H} = \frac{r_{\pm} - r_{\mp}}{4 \pi (r_\pm^2 +a^2)c_1 c_2 c_3}~.
\ee
Magentic charges are defined as $
Q^I_M = \frac{1}{4 \pi G} \int_{S^2_{\infty}} F^I =  - 2 m G^{-1} s_I c_I.$
The magnetic potentials dual to these charges can be guessed, say using the Smarr relation\footnote{A first
principle calculation of the magnetic potentials requires appropriately generalizing the formalism of
\cite{Copsey:2005se} (see also \cite{G21}) to the U(1)$^3$ theory.
Such a generalization is beyond the aspirations of the present study.}
\be
M = \frac{3}{2}\left(\frac{1}{4 G}T^+_H A^+_H + \Omega^+_\phi J_\phi\right) + \frac{1}{2} \cT (2 \pi R) + \frac{1}{2} \sum_{I=1}^{3}\Phi^I Q_M^I.
\ee
This guess is then confirmed by explicitly verifying the first law
\be
dM = \frac{1}{4G} T^+_H dA^+_H + \Omega^+_\phi d J_\phi + \sum_{I=1}^{3}\Phi^I d
Q_M^I + 2 \pi \cT   d R.
\ee
We find $\Phi_I = - \frac{\pi R s_I}{2c_I}.$ Moreover, the product
$ A^+_{H} A^-_{H}=  4 (8 \pi^2 R)^2  m^2 a^2 c_1^2 c_2^2 c_3^2 = (8 \pi G J_\phi)^2$ takes the expected form \cite{Cvetic:2010mn}.

Of particular interest is the fact that for the un-boosted solution the linear velocities $v_{z}^\pm$ \eqref{omegam} are non-zero, while the ADM momentum $P_{z}$ is zero.
Since  $v_{z}^\pm$  vanish if either $a = 0$ or any of the $\alpha_I = 0$, this
is a cumulative effect of  rotation and \emph{all three} magnetic
charges.

\subsection{Near Horizon Limit}
The near-horizon limit of the solution in section \ref{sec:m5cube} is obtained as follows. First, we write the extremal rotating solution ($m=a$)  in comoving coordinates and second we zoom in close to the horizon. More precisely, we perform
\be
r \rightarrow a+ \mu r, \quad t \rightarrow \frac{t}{\mu}, \quad \phi \rightarrow \phi + \Omega_\phi \frac{t}{\mu}, \quad z \rightarrow z + v_z \frac{t}{\mu},
\ee
with
\begin{equation}
\Omega_{\phi} = \frac{1}{2 a c_1 c_2 c_3}, \quad v_z = -\frac{s_1 s_2 s_3}{c_1 c_2 c_3},
\end{equation}
and send $\mu \rightarrow 0$. In this limit asymptotically flat region is
dispensed with. The resulting configuration is a solution of the U(1)$^3$
supergravity.  The geometry has enhanced  isometry SL$(2,
\mathbb{R})\times$U(1)$\times$U(1), as is familiar from general near-horizon
limits \cite{Astefanesei:2006dd,Kunduri:2007vf}.
The solution reads as
\begin{eqnarray}
ds^2_\rom{nh}&=& \Gamma(x)\left[-(k_\phi)^2 r^2dt^2+\frac{dr^2}{r^2} + \frac{dx^2}{1-x^2} \right] + \gamma_{\phi \phi}(x)e_\phi^2+ 2\gamma_{\phi z}(x) e_\phi \, e_z + \gamma_{zz}(x) e_z^2 \nonumber \\
A^I &=& f^I_\phi(x)e_\phi  +f^I_z(x) e_z,  \qquad h^I = h^I(x)  \label{GenExt}
\end{eqnarray}
where $e_\phi = d\phi + k_\phi r dt$, $e_z = dz + k_z r dt$. All functions are most easily expressed as $(x=\cos \theta)$
\bea
k_\phi &=& \frac{1}{2 a^2 c_1 c_2 c_3}, \quad k_z = -\frac{2 s_1 s_2 s_3}{a c_1
c_2 c_3}, \quad \Gamma(x) = \frac{1}{2} \left(\Omega_1 \Omega_2
\Omega_3\right)^{1/3}\nn \\
\gamma_{zz} &=&  \frac{4\xi}{(\Omega_1\Omega_2\Omega_3)^{2/3}}, \quad
\gamma_{z\phi} = 8 a s_1 s_2 s_3 \frac{\xi-4a^4x^2c_1^2 c_2^2c_3^2}{(\Omega_1\Omega_2\Omega_3)^{2/3}} \nn \\
\gamma_{\phi\phi} &=& \frac{ 16 a^2 s_1^2 s_2^2 s_3^2(\xi-4a^4x^2c_1^2 c_2^2c_3^2)^2+2 a^4c_1^2 c_2^2 c_3^2(1-x^2)\Omega_{1}\Omega_{2}\Omega_{3} }{\xi (\Omega_1\Omega_2\Omega_3)^{2/3}} \nn \\
f_\phi^1 &=& 4 x s_1 c_1 a^3\frac{1 + (1+2s_2^2)(1+2s_3^2)}{\Omega_{1}}, \quad f_z^1 = 4 x a^2\frac{c_1 s_2 s_3}{\Omega_{1}},\nn \\
h^I &=& (\Omega_{1}\Omega_{2}\Omega_{3})^{1/3}\Omega_{I}^{-1}
\label{NHEK}
\eea
The rest of the functions $f_\phi^2,~f_\phi^3$ and $f_z^2,~f_z^3$ are obtained by obvious cyclic permutations. In all expressions in \eqref{NHEK} the functions $\Omega_{I}$ and $\xi$ are computed at $r=a$.
An alternative presentation of these function can also be  given as in
\cite{G26}. Now let us look at various interesting limiting cases:
\begin{enumerate}
\item Upon setting all three M5 charges equal one recovers exactly the
expressions previously obtained in (11) of \cite{G26}.
\item When  M5 charges are set to zero the solution reduces to the NHEK
geometry \cite{Bardeen:1999px} times a circle, as expected.
\item  The non-trivial observation of \cite{G26}
is that in the limit of no rotation, while keep the number of M5 branes $n_I$
fixed, the solution reduces to a null orbifold of AdS$_3 \times $S$^2$,
\bea
ds^2 &=& l^2 \left( \frac{dr^2}{4r^2} - 2 r  dt dz \right) +l^2_{S_2}d\Omega_2 ,\nonumber \\
A^I &=& - \frac{n_I}{2}x d\phi, \quad z \sim z + 2\pi R,\qquad
\Phi^I =
\frac{n_I}{l}, \label{ads3geom}
\eea
where the two sphere has radius $l_{S_2} =  \frac{1}{2} \ell_p (n_1 n_2
n_3)^{1/3}$, the AdS$_3$ radius is $l = \ell_p (n_1 n_2 n_3)^{1/3} $, with
$\ell_p = (4G/\pi)^{1/3}$.
This solution has zero entropy and zero angular momentum.
\end{enumerate}

\end{document}